\documentstyle[11pt,amsfonts,amssymb]{article}

\newcommand{\be}{\begin{equation}}
\newcommand{\ee}{\end{equation}}
\newcommand{\bea}{\begin{array}}
\newcommand{\ea}{\end{array}}
\newcommand{\beqa}{\begin{eqnarray}}
\newcommand{\eeqa}{\end{eqnarray}}
\newcommand{\bean}{\begin{eqnarray*}}
\newcommand{\eean}{\end{eqnarray*}}

\def\up#1{\leavevmode \raise.16ex\hbox{#1}}

\newcommand{\gapproxeq}{\lower
 .7ex\hbox{$\;\stackrel{\textstyle >}{\sim}\;$}}
\newcommand{\lapproxeq}{\lower .7ex\hbox{$\;\stackrel
{\textstyle <}{\sim}\;$}}

\newcounter{appendice}

\def\thebibliography#1{{\bf REFERENCES\markboth
 {REFERENCES}{REFERENCES}}\list
 {[\arabic{enumi}]}{\settowidth\labelwidth{[#1]}\leftmargin\labelwidth
 \advance\leftmargin\labelsep
 \usecounter{enumi}}
 \def\newblock{\hskip .11em plus .33em minus -.07em}
 \sloppy
 \sfcode`\.=1000\relax}

\def\BI{{\rm 1\!l}}

\begin{document}

\centerline{ \LARGE BTZ Black Hole Entropy from a Chern-Simons Matrix Model}

\vskip 2cm

\centerline{A. Chaney\footnote{adchaney@crimson.ua.edu},    	Lei Lu\footnote{llv1@crimson.ua.edu} and A. Stern\footnote{astern@bama.ua.edu}   }

\vskip 1cm
\begin{center}
  { Department of Physics, University of Alabama,\\ Tuscaloosa,
Alabama 35487, USA\\}

\end{center}
\vskip 2cm

\vspace*{5mm} 

\normalsize
\centerline{\bf ABSTRACT}

We examine a Chern-Simons matrix model which we propose as a toy model for studying the quantum nature of black holes in $2+1$ gravity. Its dynamics is described by two $N\times N$ matrices, representing the two spatial coordinates.  The model possesses an internal $SU(N)$  gauge symmetry, as well as an external     rotation symmetry. The latter corresponds to the rotational isometry of the BTZ solution, and does not decouple from   $SU(N)$  gauge transformations. The system contains an  invariant which is quadratic in the spatial coordinates.  We obtain its spectrum and degeneracy, and find that the degeneracy grows exponentially in the large $N$ limit.  The usual BTZ black hole entropy formula is recovered  upon identifying the  quadratic invariant with the square of the black hole horizon radius.  The quantum  system behaves  collectively  as an integer (half-integer) spin particle for even (odd) $N$ under  $2\pi$-rotations.

\bigskip
\bigskip

\newpage
\section{Introduction}
Matrix models originating from ten-dimensional string theory have been shown in some limit to contain   geometry and gravity in less than ten dimensions.\cite{Steinacker:2010rh},\cite{Blaschke:2010ye} Most of the matrix models that have been studied, such as the IKKT model,\cite{Ishibashi:1996xs}, are of the  Yang-Mills type, with a Lagrangian which is quadratic in time derivatives. Matrix models with  Lagrangians that are first order in the time derivative are  also possible.   More specifically, they can be matrix analogues of a topological model, such as  Chern-Simons theory.\cite{Oda:1998xj},\cite{Kluson}
As has been known for some time, Chern-Simons theory allows for a description of
gravity in $2+1$ dimensions.\cite{Achucarro:1987vz},\cite{Witten:1988hc} A matrix model analogue of  Chern-Simons theory may  contain  $2+1$ dimensional geometry and gravity in some limit.  Here we  show that a Chern-Simons  matrix model    is capable of  providing a   statistical mechanical explanation of the   entropy formula for the black hole in $2+1$ gravity, i.e., the BTZ black hole.\cite{Banados:1992wn}

 Our matrix model derivation of the entropy  proceeds in a similar fashion to Carlip's derivation in \cite{Carlip:1996yb},\cite{Banados:1998ta}, which was based on the continuum Chern-Simons formulation of $2+1$ gravity.  The continuum Chern-Simons model  of \cite{Carlip:1996yb},\cite{Banados:1998ta} had physical degrees of freedom in the classical theory due to the presence of 
a boundary, the boundary being associated with the black hole horizon.\cite{Bos:1989kn},\cite{Balachandran:1994up}  These  degrees of freedom   corresponded to  edge states in the quantum theory,\cite{Balachandran:1991dw}  and the log of the degeneracy of these states  gave the entropy
\be S=\frac {\pi r_+}{2 G}\;,\label{btzntrop}\ee
where $G$ is the $2+1$ gravitational constant and $r_+$ is the outer horizon radius  of the BTZ black hole. 

 The  matrix model presented here is described in terms of two spatial coordinates,  which are represented by  $N\times N$ matrices, $\tilde X_i, \;i=1,2$.  (Time remains a continuous parameter.) Their dynamics is determined from an action which is similar to that of Chern-Simons theory on the Moyal-Weyl plane.\cite{Grandi:2000av}-\cite{Fradkin:2002qw}   Chern-Simons theory on the Moyal-Weyl plane has no dynamical content, and therefore has no hope  of describing the properties of a physical system such as a black hole.   On the other hand, the  matrix model we consider has dynamical degrees of freedom, which are analogous to the edge states of the continuum theory.  
  The system possesses an $SU(N)$  gauge symmetry, along with an additional $U(1)$ gauge symmetry. The $U(1) $ sector often plays a special role  in noncommutative gauge theories,  and that is the case here as well.   While $SU(N)$ corresponds to an internal symmetry group, the relevant $U(1)$ gauge transformations   are  external  transformations.  More specifically, they are time-dependent  rigid rotations.  The $U(1) $  rotations do not decouple from the internal $SU(N)$  transformations  in the matrix model,   and together they define a semidirect product group. We note that
rotations preserve the fundamental commutation relations of the Moyal-Weyl plane, and so rotation symmetry is also implementable for Chern-Simons theory on the Moyal-Weyl plane.  

Rigid rotation symmetry was also present in Carlip's analysis, and moreover,
it played a crucial role in the derivation of the  black hole entropy\cite{Carlip:1996yb},\cite{Banados:1998ta}.  This symmetry was associated with the isometry of the horizon.  Rotation symmetry can be utilized in a similar manner for the matrix model calculation.  As we shall show, the  physical degrees of freedom for the matrix model correspond to $N$ harmonic oscillators, which are constrained by the first class constraint generating rotations.   A unique  invariant can be written down for the model which is quadratic in the spatial coordinates $\tilde X_i$, and its spectrum and degeneracy  are easily computed. In order to make a connection with  BTZ geometry, we need to identify the quadratic invariant with a geometric invariant for the BTZ black hole which has  units of distance-squared.  A natural  choice  is $r_+^2$.  A final requirement is that we take the limit of infinite dimensional representations for  $\tilde X_i$, i.e., $N\rightarrow \infty$, for only then can we hope  to recover a two-dimensional continuous geometry from the matrix theory.  The limit of the matrix model is not  Chern-Simons theory on the Moyal-Weyl plane, and moreover the limit yields an infinite number of physical states.
Upon  taking the asymptotic limit, and identifying the quadratic invariant of the matrix model with  $r_+^2$, we obtain a degeneracy which  grows exponentially with $r_+$.   The usual formula for the BTZ black hole entropy (\ref{btzntrop}) can thus be recovered from this model.\footnote{Here we are assuming that $r_+$ is the outer horizon radius.  If one instead makes the identification with the inner horizon radius $r_-$, one recovers the results for the `exotic' BTZ black hole\cite{TownsendZhang}.}

The outline of this article is the following:
In section 2 we review the standard noncommutative Chern-Simons theory, which has no dynamical content.   In section 3 we show that physical degrees of freedom survive in a $N\times N$ matrix model analogue of the theory.  The rotation symmetry is introduced  in section 4, and a consistent invariant action is found.  The  density of states  is then computed and  found to be exponentially increasing in the large $N$ limit.  There we also show that the collective quantum system behaves as an integer (half-integer) spin particle for even (odd) $N$ under  a  $2\pi$-rotation. Concluding remarks and speculations are given in section 5.

\section{Noncommutative Chern-Simons theory}
\setcounter{equation}{0}
We now review standard noncommutative Chern-Simons theory.\cite{Grandi:2000av}-\cite{Fradkin:2002qw} 
The dynamical variables for   the theory are  a pair of infinite dimensional square matrices $X_i$, $i=1,2$, which have been referred to  in the literature as covariant coordinates.   We will take them to have units of distance. The Lagrangian is defined using an invariant trace 
\be L_{cs}(X_i,\dot X_i)=\frac k{2\theta_0} {\rm Tr}\Bigl (\epsilon_{ij}D_tX_i X_j -{ 2i}{\theta_0} A_0 \Bigr) \;,\label{NCcsLag1}\ee where the covariant derivative is defined by\be D_tX_i =\dot X_i+[A_0,X_i]\;, \ee
  and the dot denotes differentiation in the time $t$, which is assumed to be continuous. $k$ and $\theta_0$ are real constants.  The former, which we assume to be positive, is known as the level, and here takes  integer values.\cite{lq},\cite{Bak:2001ze}.  Level quantization was a result of the fact that the Lagrangian is not invariant under gauge transformations, but rather changes by a time derivative. $\theta_0$ is the noncommutativity parameter, and  has units of length-squared. $k$ and $\theta_0$ will play different roles in the subsequent sections.

$ A_0$ is an infinite dimensional square matrix whose elements correspond to  Lagrange multipliers. 
Reality for the Lagrangian requires   $A_0$ to be antihermitean, while  $X_i$ can be hermitean or antihermitean.  Our convention will be to take $X_i$  antihermitean.
 The equations of motion obtained from varying  $ A_0$ and $X_i$ are
\beqa [X_i,X_j]&=& i{\theta_0}\epsilon_{ij}\BI\label{fij0}\\D_tX_i&=&0\label{covX}\;,\eeqa 
respectively, $\BI$ being the identity.   The equation of motion (\ref{fij0}) is the Heisenberg algebra, which implies that the space spanned by coordinates $X_i$ is the Moyal-Weyl plane, with  noncommutativity parameter $\theta_0$.

The action $\int dt L_{cs}(X_i,\dot X_i)$ is invariant under noncommutative gauge transformations, where $X_i$ is in the adjoint representation.  Infinitesimal variations are of the form
\beqa \delta_\Lambda X_i&=&[X_i,\Lambda]  \cr  \delta_\Lambda A_0&=& D_t\Lambda\;,\label{ncft}\eeqa
where $\Lambda$ is an infinite dimensional  square matrix, with   time-dependent matrix elements.  The reality conditions for $X_i$ and $A_0$ are preserved provided $\Lambda$ is antihermitean.  Gauge transformations are generated by (\ref{fij0}) in the Hamiltonian formulation of the theory.  There they correspond to first class constraints, and since there is one first class constraint for every pair of matrix elements in $X_1$ and $X_2$,  no physical degrees of freedom remain in this system. 

\section{ Matrix Chern-Simons theory}
\setcounter{equation}{0}
Here we consider a finite matrix analogue of the above system.  For this let  $X_i$ and   $ A_0$ now represent finite  $N\times N$  antihermitean matrices, and let  Tr be the standard matrix trace.
A modification of the Lagrangian  (\ref{NCcsLag1}) is required in this case.
This is evident from the equation of motion
  (\ref{fij0}) which is inconsistent with the matrix trace. 
 The inconsistency is easily cured by making    $ A_0$    traceless.  
  It then  takes values in the adjoint representation of the $su(N)$ Lie algebra.   The Lagrangian in this case simplifies to 
\be L_{cs}^{(N)}(X_i,\dot X_i)=\frac k{2\theta_0} \epsilon_{ij}{\rm Tr}D_tX_i X_j  \label{NCcsLag}\ee
 Now instead of (\ref{fij0}), variations of $A_0$  lead to
 \be  [X_i,X_j]=0\label{xixj}\;,\ee  while variations in $X_i$ again give (\ref{covX}).  The equation of motion 
(\ref{xixj}) implies that the   space spanned by  spatial coordinates $X_i$ is {\it  commutative},  as opposed to what one gets from (\ref{fij0}). (Here $\theta_0$ no longer plays the role of a noncommutativity parameter.) Commuting configurations did not play a role in a derivation of four dimensional gravity from matrix models.\cite{Steinacker:2010rh} The reason was that they do not support propagating degrees  of freedom.  On the other hand,  there are no propagating degrees of freedom in a $2+1$ gravity theory. As we desire  $2+1$ gravity to emerge from the matrix model in some limit,  it is  reasonable to consider commuting configurations here.

 The Lagrangian  (\ref{NCcsLag}) possesses an $SU(N)$ gauge symmetry, with infinitesimal variations given by (\ref{ncft}).   Here $\Lambda$ are {\it traceless} antihermitean matrices.  
(The Lagrangian will be modified in the following section in order to include an additional $U(1)$ gauge symmetry.  The  additional symmetry is coupled to the $SU(N)$ symmetry in a non trivial way.)

Note that because the Lagrangian  (\ref{NCcsLag}) does not contain the previous Tr$A_0$ term, it is invariant under $SU(N)$ gauge transformations, as opposed to changing by a total time derivative.  This implies
that the constant {\it $k$ does not get quantized in this model}.    Since $X_i$ has units of length, all we require is that $k/\theta_0$ has units of inverse length-squared. These statements will also apply in section four. At the end of that section, we shall argue that $k/\theta_0$ is proportional to one over the square of the gravitational constant in  $2+1$ dimensions.

 The Poisson structure resulting from  Lagrangian  (\ref{NCcsLag}) is   given by
\be \{(X_i)_{\alpha\beta},(X_j)_{\gamma\delta}\}=\frac{\theta_0}k \epsilon_{ij}\delta_{\alpha\delta}\delta_{\beta\gamma} \;,\label{.06} \ee
where $\alpha,\beta,\gamma,\delta,...=1,...,N$ are the matrix indices.  Here (\ref{xixj}) correspond to first class constraints, with the  $SU(N)$ gauge transformations  generated from  
\be G(\Lambda)= -\frac k{2\theta_0}\epsilon_{ij}{\rm Tr}\Lambda[X_i,X_j]  \label{sunggens}\ee  This is
since $\{X_i,G(\Lambda)\}=[X_i,\Lambda]$.  Using (\ref{.06}), they form a closed  algebra 
\be  \{G(\Lambda), G(\Lambda')\}=G([\Lambda',\Lambda]) \;\label{brktGG}\ee 
 There are a total of $N^2-1$ first class constraints, which means that at least two independent physical degrees of freedom are present in the $N\times N$ matrices $X_1$ and $X_2$.  Actually, there are more.  To count the number of physical degrees of freedom, one starts with the unconstrained $2N^2-$dimensional phase space  spanned by the two matrices  $ X_i,\;i=1,2$.  The traceless parts of these matrices, call them  $ X_i^{tl},\;i=1,2$, can be taken to be elements of the $su(N)$ Lie algebra. Using the $SU(N)$ gauge symmetry, one of them, say  $ X_1^{tl}$, can be rotated to the  $(N-1)$-dimensional Cartan sub-algebra. (The result is unique up to  Weyl reflections.) This corresponds to a gauge fixing.  (Actually, it is only a partial gauge fixing, as the rotated   $ X_1^{tl}$ are invariant under rotations by the Cartan generators.)  From the gauge constraints,  the remaining matrix   $ X_2^{tl}$ must commute with the gauge fixed   $ X_1^{tl}$.  If the latter spans all of the $su(N)$ Cartan-subalgebra (we call this the generic case), then   $ X_2^{tl}$ must also be in the  Cartan-subalgebra.  So $2(N-1)$ phase space variables remain amongst   $ X_i^{tl},\;i=1,2$, after eliminating the gauge degrees of freedom.   Upon including the $SU(N)$ invariant traces of  $ X_1$ and  $X_2$, one then ends up with  $2N$ independent degrees of freedom.   They can be expressed in terms of the $SU(N)$ invariants  Tr$X_1^nX_2^m$, $n$ and $m$ being integers.  The above argument shows that only $2N$ of them are independent.
For the example of $N=2$, we can take them to be 
 \be  {\rm Tr}X_1\;,\quad {\rm Tr} X_2\;, \quad  {\rm Tr}X_1^2,\quad {\rm and} \quad {\rm Tr}X_2^2\label{4ndpndntdof} \ee More generally, (\ref{4ndpndntdof}) correspond to a minimal set of independent degrees of freedom for the matrix model.

Let us examine the simplest case of $N=2$. ($N>2$ will be studied in detail in the following section.) The  $2\times 2$ antihermitean matrices  $X_1$ and $X_2$ can be expressed as 
\be X_1=\sqrt{\frac {\theta_0}{2 k}}\; p_\mu\tau_\mu \qquad X_2=\sqrt{\frac {\theta_0}{2 k}}\; q_\mu\tau_\mu \;,\qquad \mu,\nu,... =0,...,3\;, \ee where $\tau_0=i\BI$ and $\tau_{1,2,3}=i\sigma_{1,2,3}$. $\BI$ and $\sigma_{1,2,3}$, respectively, denote the unit matrix and Pauli matrices.  Then (\ref{.06}) correspond to canonical  brackets for $q_\mu$ and $p_\mu$,
\be\{q_\mu,p_\nu\}=\delta_{\mu\nu}\label{euclid}\ee
The traces of $X_i$, which are proportional  to
 $q_0$ and $p_0$, are $SU(2)$ invariants. The traceless  parts of $X_i$, corresponding to $\vec q=(q_1,q_2,q_3)$ and $\vec p=(p_1,p_2,p_3)$, transform as vectors, so additional $SU(2)$ invariants are $\vec q^2$, $\vec p^2$ and $\vec q\cdot \vec p$, the dot denoting the scalar product.  These invariants are not all independent since the constraint (\ref {xixj}) means that the cross product of $\vec q$ and $\vec p$ vanishes.  Excluding the special (non generic) cases where one of the vectors vanishes and the other is arbitrary, we get that $\vec q$ and $\vec p$ are parallel.   Then there are a total of four independent gauge invariant quantities, $q_0$,  $p_0$,  $\vec q^2$ and $\vec p^2$, or equivalently, (\ref{4ndpndntdof}).  

\section{$ {\rm Diff}_0$ Invariant Matrix Model}
\setcounter{equation}{0}

Here we  modify the above matrix model so that it contains an additional $U(1)$ gauge symmetry. Rather than behaving like another internal gauge symmetry, the $U(1)$ transformation acts
 on the spatial indices of the coordinates $X_i$, and hence is an external symmetry transformation.  More specifically it is the analogue of  rigid rotations, which we denote  by $ {\rm Diff}_0$.    Physically, this is added in order to account for the rotational symmetry of the BTZ solution. The rigid rotation symmetry  played a crucial role in   Carlip's derivation of the  black hole entropy\cite{Carlip:1996yb},\cite{Banados:1998ta}, and we show that  it plays an important role in the analogous derivation
 for the matrix model.  After first writing down a consistent Lagrangian, we compute the spectrum of a unique invariant of the model, which is quadratic in the spatial coordinates. The entropy is obtained from the degeneracy of eigenvalues.

\subsection{Invariant Action}

We define transformations of the matrices $X_i$ in an analogous fashion to how  rotations  act on  components of a vector field $v_i,\;i=1,2$, defined on $ {\mathbb{R}}^2$.  For the latter,  infinitesimal variations  are of the form  
\be\delta_{\epsilon}  v_i  =\epsilon(t)\,({\tt L} v_ i + \epsilon_{ij} v_j) \label{ledrvA}\;,\ee 
where ${\tt L}=\epsilon_{ij}x_i \frac{\partial}{\partial x_j}$ is the angular momentum operator,  $\epsilon(t) $ is an infinitesimal time-dependent angle and $x_i$ are Cartesian coordinates  on  ${\mathbb{R}}^2$.  In analogy to this, we  write down infinitesimal  variations of  the matrices  $X_i$  of the  form
\be\delta_\epsilon X_i=\epsilon(t)({\tt L}_\Delta X_i +\epsilon_{ij}X_j) \;,\label{vrtnsetaz}\ee where   ${\tt L}_\Delta$ denotes some derivation.  We define it by ${\tt L}_\Delta M=[\Delta,M]$,  
when acting on any  $N\times N$ matrix $M$, where $\Delta$  is some time-independent $N\times N$  antihermitean matrix.    It follows from (\ref{vrtnsetaz})  that $\delta_\epsilon [X_i,X_j] =\epsilon(t){\tt L}_\Delta[X_i,X_j]$.   We need to  define the corresponding variation of $A_0$.  We take it to have the form
\be\delta_\epsilon A_0=\epsilon(t){\tt L}_\Delta A_0 +\dot \epsilon(t)\Upsilon \label{dfona0}\;\ee Since  $A_0$  is a traceless  $N\times N$ antihermitean matrix, the same must be true for $\Upsilon$.  From (\ref{vrtnsetaz})  and (\ref{dfona0}) we get the following variation of the Lagrangian (\ref{NCcsLag}) 
\be \delta_\epsilon L_{cs}^{(N)}(X_i,\dot X_i)=\dot \epsilon(t)\frac k{2\theta_0}{\rm Tr}\Bigl(\epsilon_{ij}({\tt L}_\Delta X_i) X_j+X_iX_i +\epsilon_{ij}[X_i,X_j]\Upsilon \Bigr)\label{grnvrtnlcs}
\ee
It vanishes if we set $\Upsilon=-\Delta$ and constrain Tr$X_iX_i$ to zero.  In this case,   we need to require that Tr$\Delta$=0, while the constraint  Tr$X_iX_i=0$ can be  ensured by adding
a Lagrange multiplier term to (\ref{NCcsLag}).

  More generally, there is a one-parameter family of $\Upsilon$'s for which  (\ref{vrtnsetaz})  and (\ref{dfona0})  are symmetry transformations.  It is $\Upsilon=iaX_iX_i-\Delta$,  along with the  constraint
\be {\rm Tr}(X_iX_i+i\Delta/a) =0\;,\label{cntonxsq}\ee where $a$ is  real.  The constraint can  be imposed by adding a Lagrange multiplier term to the Lagrangian.  Now the variation (\ref{grnvrtnlcs}) is a time derivative.  Using (\ref{cntonxsq}), it is $ \delta_\epsilon L_{cs}^{(N)}(X_i,\dot X_i)=\dot \epsilon(t)\frac k{2\theta_0a}(-i{\rm Tr}\Delta )$.  (Recall that $\Delta$ is antihermitean, and so its trace is imaginary. Also, for $a\ne 0$ we no longer  need to require that $\Delta$  is traceless, since Tr$\Upsilon=0$  follows from the constraint.)  The result can be extended to finite rotations.  For a $2\pi$-rotation, the corresponding action $S^{(N)}$ changes by
\be \frac {\pi k}{\theta_0a}(-i{\rm Tr}\Delta )\label{chngnactn}\ee
 We show later that its value gets fixed in the quantum theory.

 In conclusion, the action $S^{(N)}=\int dt L_{cs}^{'(N)}(X_i,\dot X_i)$, with
\be  L_{cs}^{'(N)}(X_i,\dot X_i)=\frac k{2\theta_0} \epsilon_{ij}{\rm Tr}D_tX_i X_j +\mu{\rm Tr}(X_iX_i+i\Delta/a)  \label{dfNCcsLag}\;,\ee 
is invariant under infinitesimal
variations  (\ref{vrtnsetaz}) and
\beqa\delta_\epsilon A_0&=&\epsilon(t){\tt L}_\Delta A_0 +\dot \epsilon(t)(iaX_iX_i-\Delta)\cr &&\cr  \delta_\epsilon\mu&=&-\dot \epsilon(t)\frac k{2\theta_0}  \label{difofa0mu}\;,\eeqa  where
$\mu$ is the Lagrange multiplier.  We define  (\ref{vrtnsetaz}) and (\ref{difofa0mu}) to be the infinitesimal   $ {\rm Diff}_0$ variations for the matrix model.
 (For the special case $a=0$, we should drop the term  $i{\rm Tr}(\Delta/a)$ from the Lagrange constraint and assume that $\Delta$ is traceless.)  Of course, in addition to  the  $ {\rm Diff}_0$ symmetry, the Lagrangian (\ref{dfNCcsLag}) is  invariant under  $SU(N)$ gauge transformations, where the infinitesimal variations  are (\ref{ncft}). 

The equations of motion following from the Lagrangian (\ref{dfNCcsLag})  are 
\be D_tX_i+\frac {2\theta_0}k\mu\epsilon_{ij}X_j=0\;,\label{DxeqepX}\ee
 (\ref{cntonxsq}) and (\ref{xixj}). Eq. (\ref{DxeqepX}) replaces  (\ref{covX}),
 while the condition (\ref{cntonxsq}) is new and has nontrivial consequences.  Upon restricting  $i$Tr$\Delta /a>0$,  it states that all matrix elements of $X_i$ lie on the surface of a  $2N^2-1$ dimensional sphere.  (Recall that $X_i$ are antihermitean.)  However,  from (\ref{.06}), one does not have the Poisson structure on a sphere.
 The constraint  (\ref{cntonxsq}) implies that all matrix elements  have a finite range, corresponding to the diameter of the sphere.  This means that boundary conditions must be imposed in all directions in the phase space, making quantization problematic.  [The situation is even worse for the case Tr$\Delta=0$, since then the constraint (\ref{cntonxsq}) says that all matrix elements of the antihermitean matrices $X_i$ vanish!]
This obstacle to quantization  can be easily rectified  by a simple modification of the reality conditions on the matrices $X_i$, as we describe below. 
 
\subsection{{\tt Alternative Reality} conditions}

An interesting feature of the above matrix model is that one can choose independent reality conditions for the trace and traceless parts of the dynamical matrices.   Here we exploit this feature in order to obtain a consistent quantization. More specifically, we replace the antihermitean matrices $X_i$ in  the Lagrangian (\ref{dfNCcsLag}), by matrices  $\tilde X_i$, for which

 a) the trace is real and

 b) the traceless part is antihermitean.

\noindent
 This choice is consistent with the reality of $  L_{cs}^{'(N)}(\tilde X_i,\dot {\tilde  X_i})$.  It is also consistent with  the  $SU(N)$ and $ {\rm Diff}_0$ symmetry transformations.  Infinitesimal variations for the former are given by (\ref{ncft}), while they are given by (\ref{vrtnsetaz}) and (\ref{difofa0mu}) for the latter.  We again assume that $\Lambda$ and $\Delta$ are antihermitean matrices.  $\Lambda$ is time-dependent and traceless, while $\Delta$ is a constant matrix.   From conditions a) and b),  the constraint (\ref{cntonxsq}) [with $X_i$ replaced by $\tilde X_i$] now defines a $2N^2-1$ dimensional {\it unbounded} surface. 
 
 Of course, most of the matrix elements in $\tilde X_i$ are not physical degrees of freedom.   In addition to containing the $SU(N)$ gauge degrees of freedom discussed in the  previous section, the matrix elements have a  $ {\rm Diff}_0$ gauge  degree of freedom.  In the Hamiltonian formalism, the  $SU(N)$ gauge symmetry is generated by (\ref{sunggens}) [with $X_i$ replaced by $\tilde X_i$], while the $ {\rm Diff}_0$  symmetry is generated by the first class constraint
 \be V_\Delta= \frac k{2\theta_0} {\rm Tr} \Bigl(\epsilon_{ij}( {\tt L}_\Delta \tilde X_i)\tilde  X_j + \tilde  X_i\tilde X_i+ {i\Delta}/{a}\Bigr) \approx 0\label{genofrr}\ee 
Using (\ref{.06}), one gets 
$\{\tilde X_i,V_\Delta\}={\tt L}_\Delta \tilde X_i +\epsilon_{ij} \tilde X_j$, which means that   (\ref{vrtnsetaz}) can be generated in the Hamiltonian formalism.   From \be 
\{V_\Delta, G(\Lambda)\}=G([\Delta,\Lambda])\;, \label{pbgv}\ee
and (\ref{brktGG}), the  $SU(N)$  generators $G(\Lambda)$,  along with 
the $ {\rm Diff}_0$ generator $V_\Delta$, form a closed algebra, and  yield a total of $N^2$ first class constraints in the Hamiltonian formalism.  (\ref{pbgv}) implies that  external  rotations are coupled to  the internal $SU(N)$ gauge transformations, and that the combination of the two transformations defines the action of a semidirect product group,  $SU(N)\rtimes  {\rm Diff}_0$.

 Even though there are now $N^2$ first class constraints, they do not eliminate all physical degrees of freedom from the two $N\times N$ matrices $\tilde X_1$ and $\tilde X_2$.  Following the discussion after (\ref{brktGG}), $2N$ independent degrees of freedom remain in the generic case after eliminating the $SU(N)$ gauge degrees of freedom. The $SU(N)$  invariants (\ref{4ndpndntdof}) represent a minimum set of such degrees of freedom. The physical phase space dimension reduces to $2(N-1)$ once one  introduces the additional  ${\rm Diff}_0$ gauge symmetry. (We shall construct the variables spanning the reduced   phase space explicitly in subsections  4.3.1 and 4.3.2.)  Then for the  example of $N=2$, only two of the  four $SU(N)$ invariants  (\ref{4ndpndntdof}) can be  independent  physical degrees of freedom.  More generally, a minimum of two physical degrees of freedom occur for  this matrix model.  One such degree of freedom is the
 $SU(N)\rtimes  {\rm Diff}_0$  invariant 
\be \hat{\cal I}^{(2)}= \frac 1N\Bigl(({\rm Tr}\,\tilde X_1)^2 +({\rm Tr}\,\tilde X_2)^2 \Bigr)\label{.6}\;\ee  
  The factor of $1/N$ was introduced in order to give it a universal (i.e., $N-$independent) spectrum in the quantum theory.  (\ref{.6}) is the unique quadratic invariant for the matrix model and it has units of distance-squared.\footnote{
Another quadratic  $SU(N)\rtimes  {\rm Diff}_0$  invariant is   $ {\rm Tr}( \tilde X_1^2 +\tilde X_2^2)$,  however it is constrained by (\ref{cntonxsq}) (with $X_i$ replaced by $\tilde X_i$), and hence it is not a physical degree of freedom.} For the BTZ black hole, the natural invariant with  units of distance-squared    is the square of the horizon radius.   We will identify these two invariants at the end of this section.

The spectrum of the operator analogue of  (\ref{.6}) is that of  the energy of a harmonic oscillator.
For this we note that
 the  $SU(N)$ invariants 
  Tr$\tilde X_1$ and Tr$\tilde X_2$, 
obey the Heisenberg algebra
\be \{ {\rm Tr}\tilde X_1, {\rm Tr} \tilde X_2\}=\frac {\theta_0 N}k \label{trx1trx2} \ee
This algebra  persists after  eliminating the   $ {\rm Diff}_0$ gauge degree of freedom.  For this we can impose  a gauge fixing condition.  A convenient choice is
\be \psi = {\rm Tr}\tilde X_2^2-\frac 1N( {\rm Tr}\tilde  X_2)^2\approx 0 \;,\label{gnrlgfx} \ee which along with   $V_\Delta$ form a second class set of constraints.  
 $\psi$ has zero  bracket with both $ {\rm Tr}\tilde X_1$ and $ {\rm Tr}\tilde X_2$, and as a result, the Dirac bracket of $ {\rm Tr}\tilde X_1$ with $ {\rm Tr} \tilde X_2$ is identical to (\ref{trx1trx2}).\footnote{More generally, the Dirac bracket of phase space variables $A$ and $B$ is given by $$\{A,B\}_{ {\tt DB}}= \{A,B\}+\frac{\{A,V_\Delta\}\{\psi,B\} -\{B,V_\Delta\}\{\psi,A\}  }{\{V_\Delta,\psi\}}$$}

In the quantum theory,   $ {\rm Tr}\tilde X_1$ and $ {\rm Tr}\tilde X_2$ are promoted to hermitean operators, which we denote by $\widehat{ {\rm Tr}X_1}$ and $\widehat{ {\rm Tr} X_2}$, respectively.  They  satisfy commutation relations
\be  [\widehat{ {\rm Tr}X_1},\widehat{ {\rm Tr} X_2}]=i\frac {\theta_0 N}k \label{2.29}  \ee  Raising and lowering operators, $a^\dagger$ and $a$ satisfying $[a,a^\dagger]=1$, can be introduced by writing $\widehat{ {\rm Tr}X_1}=\sqrt{\frac{\theta_0 N}{2k}}(a^\dagger+a)$ and  $\widehat{ {\rm Tr}X_2}=i\sqrt{\frac{\theta_0 N}{2k}}(a^\dagger -a)$. 
  Then the operator analogue of the invariant   (\ref{.6})  can be expressed in terms of a  number operator $a^\dagger a$, and  has the eigenvalues:
\be {\cal I}^{(2)}_n=\frac  {2\theta_0} k\Bigl ( n +\frac 12\Bigr) \;,\quad n=0,1,2,...\; \label{nrgignvlu}\ee

\subsection{Degeneracy}

We now determine the degeneracy of the eigenvalues $ {\cal I}^{(2)}_n $. We first   show that all eigenvalues are nondegenerate for the case $N=2$ in subsection 4.3.1, and then compute the degeneracy for $N>2$ in  subsection 4.3.2.

\subsubsection{$N=2$}
It is easy to see that all eigenvalues  $ {\cal I}^{(2)}_n $  are nondegenerate for  $N=2$.  For this it is convenient  to expand $\tilde X_1$ and $\tilde X_2$  in terms of $2\times 2$ matrices $\tilde\tau_0=\BI$ and $\tilde\tau_{1,2,3}=i\sigma_{1,2,3}$ according to
\be \tilde X_1=\sqrt{\frac {\theta_0}{2 k}}\;  p_\mu\tilde\tau_\mu \qquad \tilde X_2=\sqrt{\frac {\theta_0}{2 k}}\;  q_\mu\tilde \tau_\mu \qquad \ee   
In contrast to (\ref{euclid}),  $q_\mu$ and $p_\mu$ now satisfy  brackets
\be\{q_\mu,p_\nu\}=\eta_{\mu\nu}\;,\label{su2pbs}\ee  where $\eta$ is the Minkowski metric tensor $ \eta={\rm diag}(-1,1,1,1)$.\footnote{The Minkowski signature is a result of the choice of reality conditions made on the coordinates $\tilde X_i$ in  the previous subsection.  This is in contrast to the Euclidean signature that resulted from the antihermeitian coordinates $X_i$, as was seen in (\ref{euclid}).} As noted at the end of section 3, there are four independent rotationally invariant quantities $q_0$,  $p_0$,  $\vec q^2$ and $\vec p^2$, i.e., (\ref{4ndpndntdof}).   Here they generally contain  a $ {\rm Diff}_0$ gauge degree of freedom, where from (\ref{vrtnsetaz}), infinitesimal $ {\rm Diff}_0$ variations are of the form
\beqa \delta_\epsilon q_0=-\epsilon(t) p_0 &\qquad & \delta_\epsilon p_0=\epsilon(t) q_0 \cr
 \delta_\epsilon \vec q^2=-2\epsilon(t) \vec q\cdot\vec p &\qquad &  \delta_\epsilon \vec p^2=2\epsilon(t) \vec q\cdot\vec p
\eeqa
Furthermore, from  (\ref{cntonxsq}), the four rotationally invariant quantities are (weakly) constrained by 
\be q^2_0+p^2_0\approx\vec q^2+\vec p^2 +d_0\label{Neq2cnstrnt}  \;, \ee
where \be  d_0=\frac k{\theta_0 a}(-i{\rm Tr} \Delta)\label{dzero}\ee  (Again recall that ${\rm Tr} \Delta$ is imaginary.)
So here the physical phase space is two dimensional.    We can  eliminate the   $ {\rm Diff}_0$ gauge degree of freedom by imposing the gauge fixing condition $\vec q^2\approx 0$ [i.e.,  (\ref{gnrlgfx})], and furthermore solve for $\vec p^2$ using (\ref{Neq2cnstrnt}).   The remaining  independent coordinates are then $q_0$ and $p_0$, i.e.,  $ {\rm Tr}\tilde X_1$ and $ {\rm Tr}\tilde X_2$, and their Dirac bracket is identical to the  bracket $\{q_0,p_0\}=-1$.  The  rotational invariant quantity  $q^2_0+p^2_0\propto {\cal I}^{(2)} $ has the form of a harmonic oscillator Hamiltonian and its eigenvalues in the quantum theory   are $2n+1$, $n=0,1,2,...$ .  Each eigenvalue is associated with a single harmonic oscillator state. 

In an alternative quantization, one can first eliminate two of the $SU(2)$ gauge degrees of freedom (up to a $\pi-$rotation) by requiring one vector, say $\vec p$, to point along the third-direction, i.e.,  we impose  the gauge  conditions $p_1=p_2=0$.  Upon restricting to the  generic solution, $\vec p\parallel \vec q$, of the equation of motion $ [\tilde X_1,\tilde X_2]=0$, we also have that $q_1=q_2=0$.\footnote{The special solutions where one vector (either $\vec q$ or $\vec p$) vanishes, while the other is arbitrary, cannot give a discrete spectrum for the invariant  $q^2_0+p^2_0 $, using (\ref{Neq2cnstrnt}), and it is therefore inconsistent with the above result, and also (\ref{nrgignvlu}).}    The remaining nonvanishing degrees of freedom   are $q_0,p_0,q_3$ and $p_3$.  They are subject to the constraint $q^2_0+p^2_0\approx\ q_3^2+\vec p_3^2+d_0$.  While they are invariant under the remaining gauge transformations in the $U(1)$ subgroup of $ SU(2)$, they   contain the $ {\rm Diff}_0$ gauge degree of freedom. So again we find two  independent physical variables.   Now instead of taking them to be $q_0$ and  $p_0$, as we did previously, let us choose them  to be $q_3$ and $p_3$.  We can eliminate the $ {\rm Diff}_0$ gauge degree of freedom (up to a $\pi-$rotation) by imposing the constraint   $q_0\approx 0$.  Then the Dirac bracket of $q_3$ with $p_3$ is identical to the bracket $\{q_3,p_3\}=1$, and  $ q_3^2+ p^2_3$ defines another harmonic oscillator Hamiltonian.  It has  eigenvalues  $2n+1$, $n=0,1,2,...$,  in the quantum theory.   This spectrum is  identical to what we previously obtained  for the operator analogue of $q^2_0+p^2_0$, which here is weakly equal to $ p^2_0$.  In order  to make these results  consistent with the constraint  $q^2_0+p^2_0\approx\ q_3^2+\vec p_3^2+d_0$, we must have $d_0=0$, which from (\ref{dzero}) implies that $\Delta$ is traceless.     This result only applies for $N=2$.  We shall show that $\Delta$ has   nonvanishing trace when $N>2$.

In (\ref{chngnactn}) we wrote down the change of the action $S^{(N)}$ under a $2\pi$-rotation.  Using (\ref{dzero}), it is just $\pi d_0$.  Since we have found that $d_0=0$, we here get that the action is invariant under  $2\pi$-rotations.  This result is only valid for $N=2$.
 For general $N\times N$ matrices, $d_0$ depends on  $N$, as we show below, and this leads to nontrivial transformation properties of the action.  
 
\subsubsection{$N>2$}

For  $N>2$ it is convenient to expand $\tilde X_1$ and $\tilde X_2$ in the Cartan-Weyl basis of $U(N)$,
\be \tilde X_1=\sqrt{\frac{\theta_0} k}\;\Bigl(\frac{p_0\BI} {\sqrt{N}}  +i\sqrt{2}p_aH_a +i p_{-\vec\alpha }E_{\vec\alpha }\Bigr)\quad\qquad \tilde X_2=\sqrt{\frac{\theta_0} k}\; \Bigl( \frac{ q_0\BI }{\sqrt{N}}+i\sqrt{2}q_aH_a + iq_{-\vec\alpha }E_{\vec\alpha }\Bigr) \;, \ee  where  $\{H_a,\;a=1,...,N-1\}$ span the Cartan subalgebra and $E_{\vec\alpha }$ are the root vectors, $\vec\alpha$ labeling  the $N(N-1)$ roots.  $\BI$ is again the identity matrix. Thus
\beqa  [H_a,H_b]&=&0\cr&&\cr
 [H_a,E_{\vec\alpha }]&=& \alpha_a E_{\vec\alpha }\cr&&\cr
[E_{\vec\alpha },E_{\vec\beta }]&=&\left\{ \matrix{\alpha_aH_a\;,& {\rm if }\quad \vec\alpha+\vec\beta=0\cr N_{\vec \alpha,\vec\beta} E_{\vec\alpha +\vec\beta}\;,&{\rm if }\quad \vec\alpha+\vec\beta\quad{\rm is }\;{\rm a }\;{\rm root}
\cr 0\;,&{\rm if }\quad \vec\alpha+\vec\beta\quad{\rm is }\;{\rm not}\;{\rm a }\;{\rm root}}\right. \;,
\eeqa
where for all non zero roots $\vec\gamma =\vec\alpha+\vec\beta$, $ N_{\vec \alpha,\vec\beta}=N_{\vec \beta,\vec\gamma}=N_{\vec\gamma ,\vec\alpha}\ne 0$.  The   representation can be chosen such that
\be  {\rm Tr}H_aH_b=\frac 12\delta_{a,b}\qquad {\rm Tr}E_{\vec\alpha }E_{\vec\beta } = \delta_{\vec \alpha+\vec \beta,0}\qquad  {\rm Tr}H_aE_{\vec\alpha }=0\ee
Then from (\ref{.06}), we recover canonical   brackets for the $q$'s and $p$'s
\beqa  
\{q_0,p_0\}&=&-1\label{4.35}\\ \{q_a,p_b\}&=&\delta_{a,b}\label{4.36}\\
 \{q_{\vec \alpha},p_{\vec \beta}\}&=& \delta_{\vec \alpha+\vec \beta,0}\eeqa
In terms of the canonical coordinates, the generators of the $SU(N)$ transformations are the first class constraints
\beqa \Phi_a&=&\sum_{\vec \alpha}\alpha_a q_{\vec \alpha} p_{-\vec \alpha}\approx 0\cr\Phi_{\vec \alpha}&=&\sqrt{2}\sum_{a} \alpha_a( q_{-\vec \alpha}p_a- p_{-\vec \alpha}q_a) +\sum_{\vec \beta\ne \vec \alpha}N_{\vec\alpha-\vec\beta ,\vec\beta}\,q_{-\vec \beta}p_{\vec \beta-\vec\alpha}\approx 0\label{sunggenqp}
\eeqa

Following the procedure outlined in  section 3,  some of the $SU(N)$ gauge freedom can be eliminated, up to  Weyl reflections, by rotating the traceless part of  one of the matrices, say $\tilde X_1$, to the $SU(N)$ Cartan sub-algebra.  (The freedom to rotate around the Cartan generators is not eliminated by this gauge fixing,  since the resulting matrix $\tilde X_1$ is  invariant under such rotations.)  More specifically, we can fix a point on the adjoint orbit of $\tilde X_1$ by imposing the gauge fixing constraints $p_{\vec\alpha} \approx 0$. Provided that this point is not restricted to  intersect certain directions, i.e., $\alpha_a p_a = 0$, 
one gets from (\ref{sunggenqp}) that all    $q_{\vec\alpha}$'s also vanish.  Thus, in this generic case, the surviving phase space variables in $\tilde X_1$ and $\tilde X_2$ lie in the direction of the $U(N)$ Cartan subalgebra.   The nonvanishing Dirac brackets of these  variables, which include $q_0$ and $p_0$, are identical to the nonvanishing    brackets (\ref{4.35}) and (\ref{4.36}).\footnote{ Dirac brackets $\{\,,\,\}_{ {\tt DB}}$ in the generic case are computed  using
$\{\Phi_{\vec\alpha}, p_{\vec\beta}\}\approx \sqrt{2}\alpha_ap_a\delta_{\vec\alpha,\vec\beta}$ and $\{\Phi_{a}, p_{\vec\beta}\}\approx 0$.  For  two functions $A$ and $B$ on phase space, one gets
$$ \{A,B\}_{ {\tt DB}}=\{A,B\} +\sum_{\vec \alpha}\frac 1{\sqrt{2}\,\alpha_ap_a}\Bigl(
\{A, \Phi_{\vec\alpha}\}\{p_{\vec\alpha},B\}-
\{B, \Phi_{\vec\alpha}\}\{p_{\vec\alpha},A\}
\Bigr)\;,$$ where the  sum is over the roots.  The parenthesis vanishes when $A$ and $B$ are taken from the set $q_0,p_0,q_a$ and $p_a$, showing that their Dirac brackets are identical to the  brackets (\ref{4.35}) and (\ref{4.36}).  Furthermore, these Dirac brackets can be extended to   include the lines in phase space along the root directions, $\alpha_ap_a=0.$ }

The $2N-$dimensional reduced phase space spanned by $q_0,p_0,q_a$ and $p_a$ are subject to one more constraint
and contain one gauge degree of freedom associated with $ {\rm Diff}_0$.  The constraint can again be written in the form (\ref{Neq2cnstrnt}), where  here $\vec q$ and $\vec p$ are $N-1$ dimensional vectors, $\vec q=(q_1,..,q_{N-1})$ and $\vec p=(p_1,..,p_{N-1})$.   So, as stated before, there are $2(N-1)$ independent physical variables.  After imposing  (\ref{Neq2cnstrnt}) and the gauge fixing constraint  $q_0\approx 0$, we can take them to be $\vec q$ and $\vec p$, thereby eliminating $q_0$ and $p_0$,
The Dirac brackets for $\vec q$ and $\vec p$, i.e.,  (\ref{4.36}), are once again preserved by this gauge fixing.
 From the constraint (\ref{Neq2cnstrnt}), $q_0^2+p_0^2\approx  p_0^2$ is now the sum of $N-1$ harmonic oscillator Hamiltonians. If we denote the  eigenvalues of  their corresponding number operators  by $n_a=0,1,...$ , then the eigenvalues for the operator analogue of $q_0^2+p_0^2$ are $2\sum_{a=1}^{N-1} n_a+ {N-1+d_0}$.  This means that the eigenvalues of the
 $SU(N)\rtimes  {\rm Diff}_0$  invariant (\ref{.6}) are 
\be\frac  {2\theta_0 } k\Bigl(\sum_{a=1}^{N-1} n_a+\frac {N-1+d_0}2 \Bigr)\label{427}\ee
  In comparing with (\ref{nrgignvlu}),  $n=\sum_{a=1}^{N-1} n_a+\frac {N+d_0}2-1$.  Since the  eigenvalues of  (\ref{nrgignvlu}) and (\ref{427})  must agree, we have that 
\be n=\sum_{a=1}^{N-1} n_a\;,\qquad \quad d_0=2-N  \label{d0frN}\ee  Thus only one value of $d_0$ is possible for any given $N$.  For $N=2$ the result is $d_0=0$, which agrees with what we found previously.

The degeneracy $g_n^{(N)}$ of the $n^{\rm th}$ excited  level of the matrix model is identical to what  one  would get from the $N-1$ dimensional isotropic harmonic oscillator.  The system can be expressed in terms of  $N-1$ pairs of raising  and lowering  operators, $\hat a_a^\dagger$ and $\hat a_a, $ respectively. Since we want  the degrees of freedom to be associated with  those of a gravitational field,  it makes sense to identify $\hat a_a^\dagger$ and $\hat a_a $ with {\it bosonic } creation and annihilation operators.  
In this picture, the  $n^{\rm th}$ excited  level  consists of  states of $n$ identical bosons occupying $N-1$ sites.  The degeneracy  $g_n^{(N)}$ is a sum of the  number $p(n,k)$ of partitions of $n$ into $k$ parts,
\be g_n^{(N)}=\sum^{N-1}_{k=1} p(n,k)\label{gsubn}\ee
This sum is known to be identical to the number of partitions $p_{N-1}(n)$ of $n$ into parts none of which exceeds $N-1$.\cite{Grosswald}
In the asymptotic limit $N,n\rightarrow\infty$, with $N\ge n$, it is given by the Hardy-Ramanujan formula 
\be g_n^{(N)}\rightarrow \frac 1{4n\sqrt{3} }\exp{\Bigl(\pi\sqrt{\frac{2n}3}\Bigr)}\label{4.32}\ee

We define the entropy as the log of the degeneracy. Upon taking the log of (\ref{4.32}) and substituting (\ref{nrgignvlu}), one gets the following result for the entropy of the $n^{\rm th}$ excited level  in the asymptotic  limit
 \be S_n\sim \pi \sqrt{\frac{k {\cal I}^{(2)}_n}{3\theta_0}}\label{ntropee}\ee 
The usual formula for the  BTZ black hole  entropy (\ref{btzntrop})  is recovered  when we make the   identification of the  quadratic invariant (\ref{.6}) with the square of the black hole horizon radius, $r_+^2$ and the identification of constants in the two theories, $k/\theta_0$ and $\frac{3}{4G^2}$. The latter sets the scale for the  eigenvalues of (\ref{.6}), and hence $r_+^2$.  It says that they are separated by $\frac 83 G^2$, and that the smallest value for the horizon radius is   $\frac 2{\sqrt{3}}\,G$. 

 Concerning the asymptotic limit, we  assumed above that   both  $N$ and $n$ go to $\infty$, with $N\ge n$.  Other   limits are possible.  The leading order entropy  doesn't grow as fast as in (\ref{ntropee}) and can depend on $N$ for those cases.  For example, if one instead  holds the size $N$ of the matrices  fixed while taking $n\rightarrow\infty$, then\cite{Grosswald} $g_n^{(N)}\rightarrow \frac{n^{N-2} }{(N-1)!(N-2)!}$.  The leading order behavior of the entropy is logarithmic in this case,  \be S_n\sim (N-2) \log \frac{k {\cal I}^{(2)}_n}{2\theta_0},\;\;\qquad N>2\ee 

Finally, we comment on the rotational properties of the collective system.
From  (\ref{chngnactn}) and  (\ref{dzero}),  the change of the action $S^{(N)}$ under a $2\pi$-rotation is  $\pi d_0$.  Our result (\ref{d0frN}) for arbitrary $N$, then  gives a change of $\pi (2-N)$.  So under a  $2\pi$-rotation, the phase $\exp{i S^{(N)}}$ picks up a factor $(-1)^N$.  This means that the collective quantum system behaves as an integer (half-integer) spin particle for even (odd) $N$ under  a  $2\pi$-rotation.

\section{Concluding remarks}

We have shown that the BTZ black hole entropy formula emerges  from a Chern-Simons matrix model in the asymptotic limit $N,n\rightarrow\infty$, with $N\ge n$.  One does not recover  Chern-Simons theory on the Moyal-Weyl  plane  in this limit, even though both are expressed in terms of two infinite dimensional matrices representing the spatial coordinates.  This is fortunate because  Chern-Simons theory on the Moyal-Weyl  plane  has no  dynamical content.  The two systems also differ by the fact that our matrix model has  only commutative configurations, which persist in the limit, while  (\ref{fij0}) states that Chern-Simons theory on the Moyal-Weyl  plane has noncommutative configurations.  An important ingredient in the matrix model is the $ {\rm Diff}_0$ symmetry.  In addition to corresponding to the rotational symmetry of the BTZ solution, it is responsible for the first class constraint (\ref{genofrr}), from which the density of states was  computed.  The entropy law followed after identifying the   invariant (\ref{.6}), which was quadratic in the coordinates $X_i$, with the square of the horizon radius, $r^2_+$, and taking the asymptotic limit.  From the identification, one  gets a harmonic oscillator  spectrum for $r^2_+$.  From a further identification of the constants of the two systems, one sets the scale of the eigenvalues of $r_+$.  For example, the  ground state value of   $r_+$ is $\frac 2{\sqrt{3}}\,G $.  An exact expression for the entropy can  be given for any  eigenvalue for  $r_+$ and for any $N$. Lastly, we found that the collective quantum system behaves as an integer (half-integer) spin particle for even (odd) $N$ under  a  $2\pi$-rotation.

  It remains to be seen whether the BTZ  geometry can be recovered from this matrix model, with perhaps some  modifications,  in some asymptotic limit.   In this regard, the  4 dimensional  Schwarzschild and  Reissner-Nordstr\"om black hole geometries were shown to emerge from a matrix model in a `semiclassical' limit.\cite{Blaschke:2010ye}  The relevant matrix model  in that case was of the Yang-Mills type, with an action that involved quadratic and  higher order terms.  It also required an embedding  in  higher dimensions.  An analogous derivation of the BTZ solution, starting from  a higher dimensional Yang-Mills type matrix model, may also be possible.  Our work suggests that the total action should include a topological  term in order to recover the correct BTZ entropy formula.   It also suggests that commuting configurations  and the Diff${}_0$ symmetry should play an important role.  A generalization of the topological action examined here can be made to any odd number of dimensions.  Questions concerning whether or not the computations carried out here are  generalizable to higher dimensions, or if topological terms play a role in higher dimensional matrix models, are worth pursuing.

\bigskip
{\Large {\bf Acknowledgments} }

\noindent
We are very grateful to A.  Pinzul for valuable discussions. 
A.S. was supported in part by the DOE,
Grant No. DE-FG02-10ER41714.

\end{document}